\documentclass{article}
\usepackage{spconf,amsmath,epsfig}
\usepackage{graphicx}
\usepackage{subcaption}
\usepackage{url}
\usepackage{hyperref}
\let\OLDthebibliography\thebibliography
\renewcommand\thebibliography[1]{
  \OLDthebibliography{#1}
  \setlength{\parskip}{0pt}
  \setlength{\itemsep}{0pt plus 0.3ex}
}

\title{Supervised Neural Networks for RFI Flagging}
\name{Kyle Harrison, Amit Kumar Mishra\thanks{Thanks to the Inter-university Institute for Data Intensive Astronomy (IDIA).}}
\address{University of Cape Town\\
Department of Electrical Engineering\\
Cape Town, South Africa}
\begin{document}
%\ninept
%
\maketitle
\begin{abstract}
Neural network (NN) based methods are applied to the detection of radio frequency interference (RFI) in post-correlation, post-calibration time/frequency data. While calibration does affect RFI for the sake of this work a reduced dataset in post-calibration is used. Two machine learning approaches for flagging real measurement data are demonstrated using the existing RFI flagging technique AOFlagger as a ground truth. It is shown that a single layer fully connect network can be trained using each time/frequency sample individually with the magnitude and phase of each polarization and Stokes visibilities as features. This method was able to predict a Boolean flag map for each baseline to a high degree of accuracy achieving a Recall of $0.69$ and Precision of $0.83$ and an F1-Score of $0.75$. 

The second approach utilizes a convolutional neural network (CNN) implemented in the U-Net architecture, shown in literature to work effectively on simulated radio data. In this work the architecture trained on real data results in a Recall, Precision and F1-Score $0.84$, $0.91$, $0.87$ respectfully.

This work seeks to investigate the application of supervised learning when trained on a ground truth from existing flagging techniques, the results of which inherently contain false positives. In order for a fair comparison to be made the data is imaged using CASA's CLEAN algorithm and the U-Net and NN's flagging results allow for $5$ and $6$ additional radio sources to be identified respectively.
\end{abstract}

\section{INTRODUCTION}
The dramatic rise in sensitivity and the increased bandwidth of modern radio telescopes has caused an ever growing increase in the spectrum overlap between astronomical measurements and man-made radio communication. These man-made signals, viewed as radio frequency interference (RFI) in these applications, are often orders of magnitude more powerful than the faint astronomical emissions.

There exist many techniques for the mitigation, detection and excision of RFI in radio astronomy. Many of these techniques take place in different stages along the data capture and processing pipeline.  Some examples of pre-correlation mitigation techniques include; governmental legislature to reduce the presence of man-made signals and separate reference antennas directed at common sources of RFI.\cite{ref_ant}
Despite attempts at reducing the presence of RFI through pre-correlation hardware techniques, post-correlation of RFI through software is required in the time/frequency and/or antenna space to ensure no corrupted data is further processed.

This paper examines an application of neural networks (NN) to detect RFI in post-correlated interferometry measurements. The NNs are trained on existing techniques for RFI flagging, a Boolean mask identifying non-astronomical time/frequency samples, in order to learn and identify the relationship between RFI and astronomical data. 

Majority of existing flagging algorithms are created to ensure as much RFI is identified as possible, their outputs often contain many false positives. This is due to the extremely detrimental effect the high magnitude RFI outliers cause in the imaging pipeline. This makes minimising over-fitting of the NN imperative as well as making comparison metrics between algorithms more complicated.

The high volumes of data produced from interferometers for each observation ensure that even with a high number of measurements flagged as corrupted it is still possible to produce high quality science images. It is still advantageous for a flagging algorithm to minimize the number of false positives in order reduce image noise, increase the number sources identified and their associated flux.

\section{EXISTING APPROACHES TO FLAGGING}
Pre-correlation techniques must handle vast amounts of data and in order to be usable must have a low complexity. While advantageous to perform an operation like blanking or subtraction of on-line short RFI bursts, any remaining RFI must be removed in the data reduction and imaging pipeline.\cite{blanking}

\subsection{AOFLAGGER}
The AOFlagger pipeline is a popular default RFI flagging application used in multiple observatories. It operates iteratively using surface fitting and thresholding techniques on a single baseline.\cite{offringa_soft} The SumThreshold technique in AOFlagger is a combinatorial thresholding method operating in the time/frequency domain.  A selected threshold $x_M$ where $M$ is the number of samples surrounding the target is iteratively decreased until the sum of the amplitudes in $M$ exceeds the threshold $x_M$ and all visibilities are flagged.\cite{aoflagger}\cite{offringa}

\subsection{CASA TF CROP}
TFCrop is a autoflag algorithm created for NRAO's CASA.\cite{casa} It attempts to detect the presence of outliers on the 2D time-frequency plane, operating on chunks of time on each baseline and correlation independently. It seeks to create a bandshape template by iteratively fitting third-degree polynomials and calculating the standard deviation between the data and the fit. The result is then divided between all timestamps in the chunk in order to calculate deviation from the mean and identify narrowband RFI. The variable threshold is used to iteratively flag RFI and the entire process is repeated in the opposite direction, averaging over time or averaging over frequency.\cite{tfcrop}

\subsection{GRIDflag}
A separate approach to the techniques operating in the 2D time-frequency plane utilizes the UV plane to detect corrupted data. The tracks each baseline forms in the UV plane are interpolated onto a regular grid in order for imaging algorithms to apply a fast Fourier transform (FFT).  Each baseline will often contribute multiple visibilties from different time intervals to a particular cell. These visibility samples will be a measure of the same celestial information - but may record different RFI owing to the time of observation. 

This distinction is how the GRIDflag algorithm was implemented. It generates RFI thresholds based on the differences between visibilities within a UV-bin.\cite{uv}

\subsection{MACHINE LEARNING TECHNIQUES}
The application of supervised machine learning for RFI flagging have recently been investigated in literature.

In the work done by \textit{Mosiane O, et al} implementations of K-Nearest Neighbour (k-NN), Random Forest Classifier (RFC) and Naive Bayesian (NB) where trained on measurements flagged using AOFlagger as a ground truth.\cite{knn}\cite{rfc}\cite{bayes} Each time-frequency baseline was flattened and concatenated together and a sliding window was used to extract statistical features. The RFC showed high predictive capabilities with an F1 score of $0.93$.\cite{ska}

Neural network techniques have primarily been treating the time/frequency domain as a semantic segmentation problem. To this extent various architectures of convolutional neural networks have been implemented with simulated time/frequency data. HIDE \& SEEK is an open source package simulating single dish radio survey data and uses a U-Net architecture for RFI detection.\cite{akeret2017radio} 

More recently work has been done to simulate Hydrogen Epoch of Reionization Array (HERA) visibility data with simulated RFI in order to act as a ground truth. A Deep Fully Convolutional Neural Network (DFCN) in the U-Net architecture is trained on a single polarization's magnitude and/or phase. Their results proved more effective than the currently used watershed algorithm in their pipeline and showed improvement in using magnitude and phase as features over just magnitude. After training prediction is done on real HERA-67 data and achieved a recall of $81\%$ and precision of $58\%$.\cite{hera_sim}

\section{METHODOLOGY: DATA PREPROCESSING}\label{sec:preproc}
Data preprocessing is a fundamental step in the machine learning process as it directly affects the ability of a model to learn. The datasets used in these investigations are from MeerKAT science observations. These are stored in the form of CASA Measurement sets (MS) and accessed using Taql, a high level SQL-like table query standard. While effective for query operations associated with astronomical applications it has a high data access overhead with a complexity incompatible with machine learning. To overcome this the storage used for rapid data access during the learning is process is accomplished using HDF5 files, specifically designed for high performance I/O processing and storage.

The data used in these application is fully polarized having undergone four complex correlations. Each correlation of the complex visibility from each baseline is in the form $X_1X_2$, $X_1Y_2,$ $Y_1X_2$, $Y_1Y_2$ for the $X$ and $Y$ polarizations for pairs of telescopes 1 and 2. 

A neural network is proposed which is trained on each time/frequency sample individually, with features being represented by the magnitude and phase of each polarization and Stokes visibilities. This accounts for a total of 16 features from 8 magnitude and 8 phase vectors.

The four polarizations are used to compute the four complex Stokes visibilties for dual linearly polarized antennas to serve as these additional features. Stokes parameters are used to describe the total intensity and the degree of polarization as another characteristic for the network to identify RFI versus non-RFI. 

\begin{equation}
        \begin{aligned}[b]
I = X_1X_2 + Y_1Y_2 \\
Q = X_1X_2 - Y_1Y_2\\
U = X_1Y_2 + Y_1X_2 \\
V = -j(X_1Y_2 - Y_1X_2)
        \end{aligned}
\label{eqn2.qo}
\end{equation}

Justification for using each polarization and Stokes visibility as additional features is shown by the investigations into the Pearson correlation coefficient, measuring the linear correlation between the magnitude of each polarization and Stokes parameter over all baselines combined. Where: $\operatorname{cov}$ is the covariance, $\sigma_X$ is the standard deviation of X, $\sigma_Y$ is the standard deviation of Y. 

\begin{equation}\label{eq:pearson}
\rho_{X,Y}= \frac{\operatorname{cov}(X,Y)}{\sigma_X \sigma_Y}
\end{equation}

\begin{figure}[!h]
\begin{subfigure}{\columnwidth}
  \centering
  % include first image
  \includegraphics[width=.8\columnwidth]{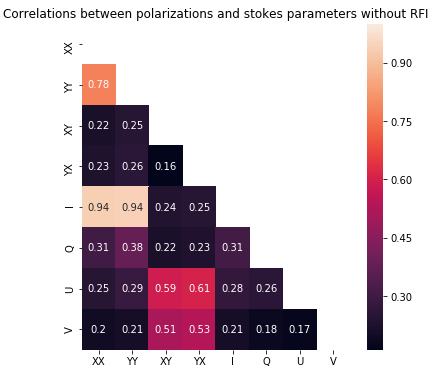}  
  \caption{Non-RFI data only}
  \label{fig:corr_norfi}
\end{subfigure}
\begin{subfigure}{\columnwidth}
  \centering
  % include second image
  \includegraphics[width=.8\columnwidth]{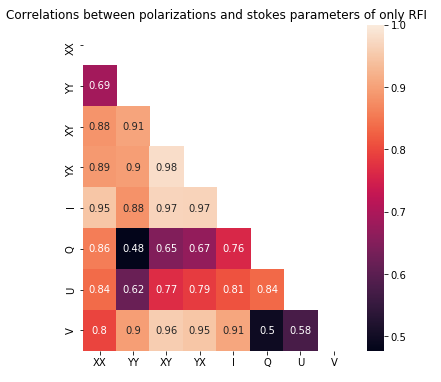}  
  \caption{RFI data only}
  \label{fig:corr_rfi}
\end{subfigure}
\caption{A diagonal heat map of correlation coefficients between each magnitude of the polarizations and Stokes parameters of the entire dataset, representing half the features. The RFI flagged by AOFlagger is removed in (a) leaving only the celestial information, while the isolated RFI is shown in (b). The figures show a low correlation between majority of the non-RFI features, while a higher correlation in the RFI.}
\label{fig:corr}
\end{figure}

Results of the correlation coefficients are shown in Figures \ref{fig:corr_norfi} and \ref{fig:corr_rfi}. These show a high correlation between all the RFI samples from all baselines flagged using AOFlagger, and conversely a low correlation between only the astronomical data.  It is assumed that by using each additional feature a subsequent intrinsic characteristic of RFI is being represented, allowing the network learn from a more complete representation of the RFI present and improve its predictive capabilities. It was shown experimentally that this is the case. With metrics measuring the performance of later networks on different combinations of features showing that using all 16 features resulted in higher accuracy. 

The 16 feature vectors comprising of the magnitudes and phases of each polarization and Stokes visibility are then normalized by a min-max scaling of the entire time/frequency spectrum in order to maintain the euclidean relationship between high magnitude RFI outliers and the astronomical data:

\begin{equation}
x' = \frac{x - \text{min}(x)}{\text{max}(x)-\text{min}(x)}
\end{equation}

The amount of time-frequency data from a single measurement set is vast. A MeerKAT observation using all 64 telescopes would account for $(64\times(64-1))/2=2016$ baselines, often with around $4000$ frequency bins and $2000$ time samples. As not all baselines are necessary for the training process, the pre-processing and storage of data in HDF5 files is done using randomly chosen baselines in order to ensure a fair sample of which telescope's data is used.

\section{NEURAL NETWORK ARCHITECTURE}
A neural network design is proposed which operates on each time-frequency sample with the 16 features derived from the magnitudes and phases of each polarization and Stokes visibility described in Section \ref{sec:preproc}. The neural network is optimized for the number of layers and nodes, and a convolutional neural network based on the U-net architecture is described.

\subsection{OPTIMIZATION OF A NEURAL NETWORK}
Ordinarily a grid search would be used to optimize the hyperparameters of the NN. The number of features combined with the number of baselines necessary to capture sufficient representations of RFI leads to high complexity. It would prove impossible to optimize the network in a complete manner with combinations of; the number of nodes, number of layers, loss functions, activation functions and optimization functions. To overcome this, a heuristic approach is taken evaluating different hyperparameters independently.

Investigations are carried out using Keras into the effect of differing the number of layers with the first hidden layer maintaining 512 nodes and each subsequent layer halving the number of nodes. Further investigations are carried out into varying the number of nodes in order to identify a baseline architecture for further fine-grained optimization. 

The \textit{Adam} optimizer is used over all tests and iterations. It is now widely used research over stochastic gradient descent (SGD), combining the benefits of the SGD extensions; Adaptive Gradient Algorithm (AdaGrad) and Root Mean Square Propagation (RMSProp) while maintaining low complexity and memory requirements. To this extent the Adam optimizer utilizes the adaptive, per-parameter learning rate of AdaGrad, which improves performance on problems with sparse and/or noisy gradients. As well as using the average of the second moments of gradients.\cite{adam}  

The activation functions of the hidden layers are fixed as ReLU, $f(x) = x^+ = \operatorname{max}(0, x)$, chosen for their properties of sparsity and reduced likelihood of a vanishing gradient. The output is a single node with a sigmoidal activation function $S(x) = \frac{1}{1 + e^{-x}}$. This produces a probabilistic prediction between $0$ and $1$ from the input vector. Setting a threshold at $0.5$ would identify everything from $<0.5$ as non-RFI and $>0.5$ as RFI. This threshold could be varied to produce an optimum output favouring RFI prediction.

The dataset is split into $70\%/30\%$ training and test data. Each iteration is trained using 5-fold cross-validation with a mean of the resulting metrics being taken. As the amount of data cannot be loaded into memory a data generator is used to extract data from the HDF5 files each time a new batch is requested. The order of each data batch is shuffled at the end of each epoch to reduce overfitting.

The number of epochs for training each iteration are constrained using an early stopping callback monitoring the F1 score. This reduces the overall time required for training but can have the effect of preventing overfitting by stopping the training process if the F1 score of the validation data has not shown improvement above $0.01$.

Evaluation of the results is done through the use of precision, recall and F1 score for binary classes. ROC and AUC are used as further metrics to evaluate the classification problem at different threshold values.

The receiver operating characteristics (ROC) curve is a metric used in binary classification, plotting the True Positive Rate against the False Positive Rate as the discrimination threshold is varied. The area under curve (AUC) describes the ability of the model to discriminate between the binary classes, where 1 would be able to perfectly distinguish between classes.

Precision is an indication of the percentage of correctly identified astronomical data, or non-RFI events. Recall describes the percentage of correctly identified non-RFI events taking into account incorrectly identified RFI. The F1 score is simply the harmonic mean of precision and recall, used as a metric for evaluating the overall predictive capabilities of a classifier. These metrics are used over accuracy, as the identification of RFI represents an imbalanced classification, where the amount of RFI is often far lower than that of non-RFI.

\begin{align}\label{eq:prec_rec_f1}
    Precision &= \frac{\text{true positive}}{\text{true positive + false positive}}\\
Recall &= \frac{\text{true positive}}{\text{true positive + false negative}}\\
F1 score &= 2 \times \frac{\text{precision} \times \text{recall}}{\text{precision} + \text{recall}}
\end{align}

\subsection{NUMBER OF LAYERS}
These tests investigate varying the number of hidden layers from one to five with the initial hidden layer containing 512 nodes and each successive layer halving the previous number of nodes. This implementation was found effective through separate trail and error experimentation, where it was evident decreasing the number of nodes each layer lead to improvements in performance over an increasing or the same number of nodes. This is expected to be from a reduction in overfitting by reducing the number of high level features extracted each layer.

\begin{figure}[h]
\begin{subfigure}{\columnwidth}
  \centering
  % include first image
  \includegraphics[width=.8\columnwidth]{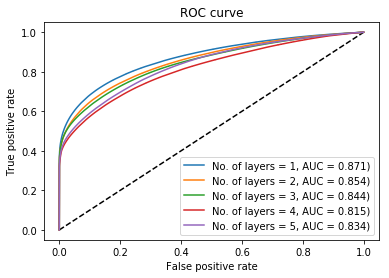}  
  \caption{Receiver operating characteristics (ROC) for a varying number of layers. }
  \label{fig:roc}
\end{subfigure}
\begin{subfigure}{\columnwidth}
  \centering
  % include second image
  \includegraphics[width=.8\columnwidth]{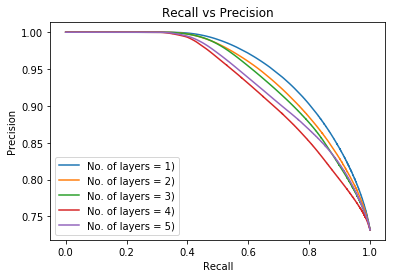}  
  \caption{Recall vs precision for a varying number of layers.}
  \label{fig:prec_rec}
\end{subfigure}
\caption{Plots showing evaluation of the results into optimizing the network for the number of layers which perform with the highest predictive capabilities. The ROC curves and related AUCs are extremely similar. This may be a demonstration of how the application of neural networks to this problem has an inherently low complexity and the non-linear discriminating hyperplane has a low dimensionality.}
\end{figure}

The mean of the 5-fold cross validation for; ROC, precision, recall and F1 score are used as metrics. These assist in identifying the performance of each network on the test data into varying the number of layers are evaluated in Figure \ref{fig:roc} and \ref{fig:prec_rec}. The ROC curve demonstrates that the differences between multiple layers is minor. A single layer with 512 nodes proves to be the most effective. This effect is likely due to the hyperplane for RFI identification not existing in a high-dimensional space, as many simple time-frequency thresholding techniques prove effective for the problem. It is probable having more layers resulted in overfitting while also increasing processing time. 

The precision recall curves for each layer demonstrate the same results as the ROC. From these results it is clear a single layer is an effective implementation going forward.

\subsection{NUMBER OF NODES}
\begin{figure}[h]
\begin{subfigure}{\columnwidth}
  \centering
  % include first image
  \includegraphics[width=.8\columnwidth]{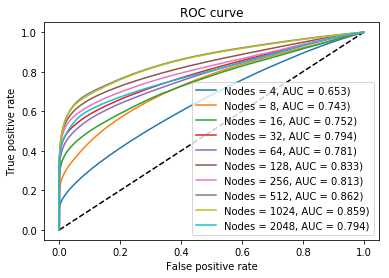}  
  \caption{Receiver operating characteristics (ROC) for a varying number of nodes. }
  \label{fig:roc_nodes}
\end{subfigure}
\begin{subfigure}{\columnwidth}
  \centering
  % include second image
  \includegraphics[width=.8\columnwidth]{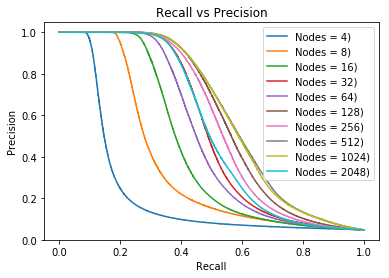}  
  \caption{Recall vs precision for a varying number of nodes.}
  \label{fig:rec_nodes}
\end{subfigure}
\caption{Plots showing evaluation of the results into optimizing the network for the number of nodes in a single layer which perform with the highest predictive capabilities. The resulting AUCs are shown to increase up until 512 nodes at which point they begin to decrease, similarly this is shown in the recall and precision curves.}
\end{figure}

In accordance with the previous tests and their control hyperparameters, an investigation into varying the number of nodes in a single layer in carried out.  The resulting ROC and precision recall curves are shown in Figure \ref{fig:roc_nodes} and \ref{fig:rec_nodes}. These plots show how after $64$ nodes the difference in accuracy is minimal. The loss of accuracy for $2048$ nodes is likely due to the training reaching the maximum number epochs at 3000 not being sufficient to reach convergence.  
For further optimization and a network with a single layer of 512 nodes is selected.  While reducing the number of nodes will aid in complexity, the network is so small already that the biggest overhead in training is likely data transfer not weight updates.

\section{U-net}
The second neural network proposed for this application is the U-net architecture, proposed by \textit{Ronneberger et al} as an extension of the convolutional neural network (CNN) used in image segmentation.\cite{unet} This architecture of CNN has been used for the prediction of RFI in related works and has shown significant results.\cite{akeret2017radio}\cite{hera_sim}

The architecture differs from a traditional CNN by using an increasing number of features in each convolutional layer as the network approaches the fully connected convolutional layer, whereby it then decreases the number of features towards the output layer. The final layer applies a $1\times1$ convolution to map the final layer in order to formulate a probabilistic decision with the use of a sigmoidal activation function. The models loss function is evaluated with binary cross entropy and the optimization algorithm used is Adam.

In order to utilize the entire time-frequency plane of each baseline the 'images' are generated by slicing the time-frequency data into $100\times100$ segments, each with all $16$ features. This non power $2$ image size requires the addition of cropping layers and a zero-padding layer before the final convolutional layers in order for the input shape to be matched after the last upsampling operation. Overfitting is attempted to be minimised through the use of a dropout layer of $0.3$ and batch normalization limiting the activation's after each double convolutional layer. 

\begin{figure}[h]
\begin{subfigure}{\columnwidth}
  \centering
  % include first image
  \includegraphics[width=.8\columnwidth]{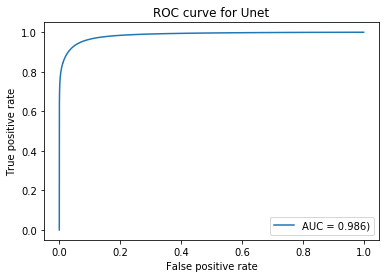}  
  \caption{Receiver operating characteristics (ROC) for U-Net. }
  \label{fig:u_roc}
\end{subfigure}
\begin{subfigure}{\columnwidth}
  \centering
  % include second image
  \includegraphics[width=.8\columnwidth]{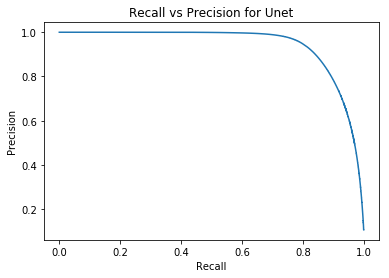}  
  \caption{Recall vs precision curve for U-Net.}
  \label{fig:u_rec}
\end{subfigure}
\caption{Plots showing the results of predictions generated by the U-Net architecture. An AUC of $0.98$ is extremely high and may be characterized by the architecture over-fitting.}
\end{figure}

Training is done using a training/test split of $70/30$ for all the images. Data transfer is handled through the use of a data generator to fetch each batch from a chunked HDF5 file. The ROC and recall precision curves in Figure \ref{fig:u_roc} and \ref{fig:u_rec} demonstrate remarkable predictive capabilities with an AUC of $0.98$. In this case where the training data contains over-flagging false positives, an accuracy that high leads to overfitting despite multiple steps taken to reduce this.

\subsection{RESULTS}
The goal of training networks on existing flagging results was to investigate whether they could identify RFI more accurately in order to prevent the existing false positives when predicting their own training data.  By generating an automated system to re-flag data and capture additional celestial information it is hoped further insights can be gained once the data is imaged.

Results are generated by predicting RFI on the entire dataset, including training data. These resulting probabilistic outputs could have their decision threshold varied per baseline in order to maximise a specific metric. So far attempts at optimising this have proved unsuccessful and a threshold is set at $Data  < 0.5 < RFI$.

Table \ref{tbl:metrics} shows the resulting metrics over the entire dataset between the two methods when using a threshold of $0.5$ compared to AOFlagger as the ground truth. It is interesting to note that when monitoring the these metrics during the training process when predicting the randomly shuffled $30\%$ test data at the end of each epoch - there is little difference. This is evidence of both networks being robust to overfitting as the results of training and testing remain within $~5\%$ of each other.

\begin{table}[h]
\centering
\caption{Evaluation metrics between the NN and U-Net}
\label{tbl:metrics}
\begin{tabular}{|l|l|l|l|}
\hline
 & \textbf{Precision} & \textbf{Recall} & \textbf{F1} \\ \hline
\textbf{Neural Network} & 0.828 & 0.6917 & 0.754 \\ \hline
\textbf{U-Net} & 0.905 & 0.837 & 0.870 \\ \hline
\end{tabular}
\end{table}

These resulting Boolean masks are copied into the CASA MS and are imaged using CASA's CLEAN algorithm. The algorithm is used in image deconvolution, iteratively working on the highest values of identified point sources and subtracting a small gain convolved with the point spread function of the observation. 

\begin{figure}[h]
\centering
\includegraphics[width=\columnwidth]{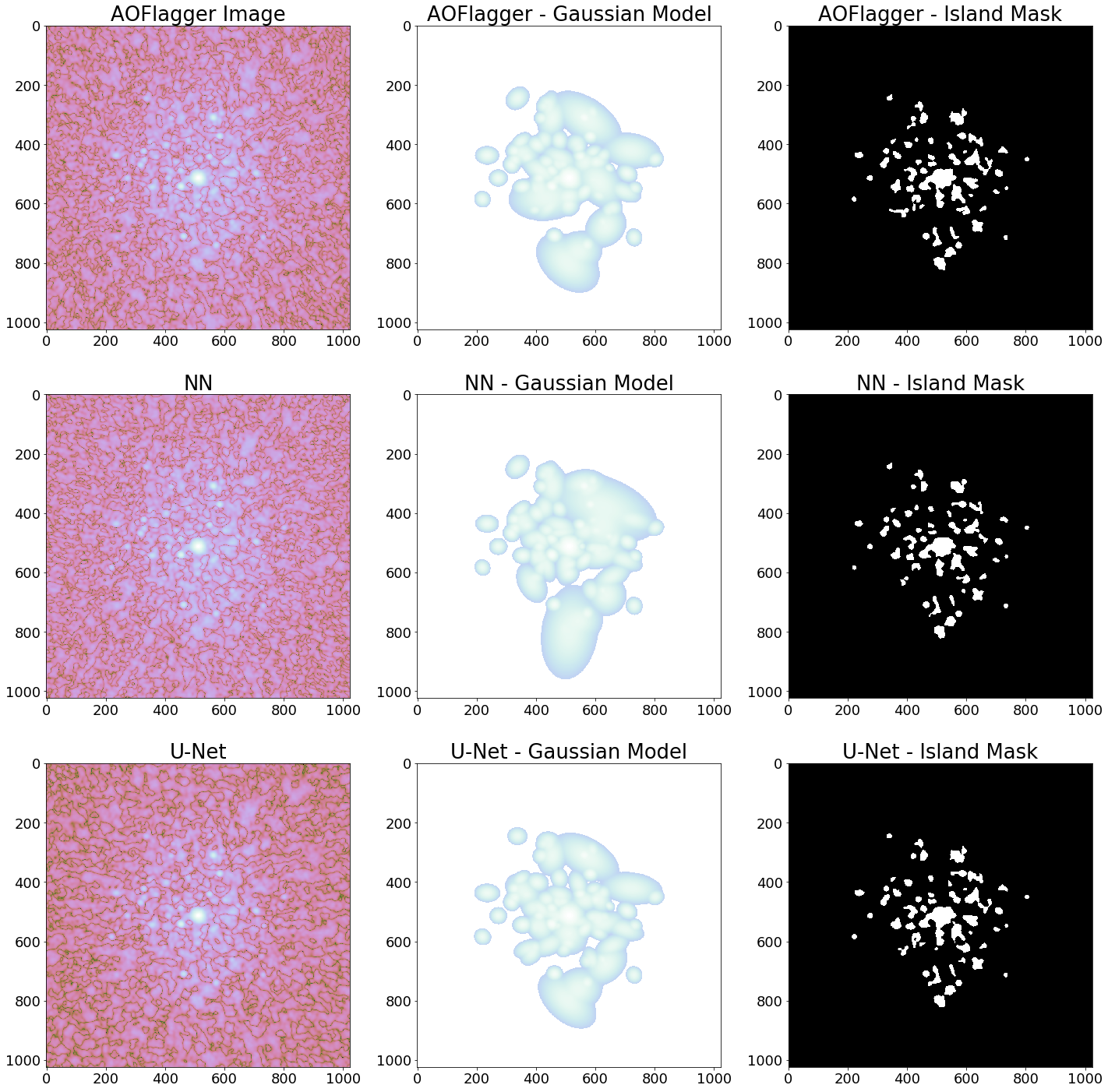}
\caption{Images generated with CASA CLEAN using each flagging method compared to PyBDSF generated Gaussian models and island masks.}
\label{fig:imgs}
\end{figure}

Python Blob Detection and Source Finding (PyBDSF) is used to generate Gaussian and island models of the identified sources in images. The images generated using flags from AOFlagger, the NN and the U-Net are processed and shown in the resulting Figure \ref{fig:imgs}. The CLEAN algorithm is known to be robust and therefore not many changes can be seen between each image.

\begin{figure}[h]
\centering
\includegraphics[width=\columnwidth]{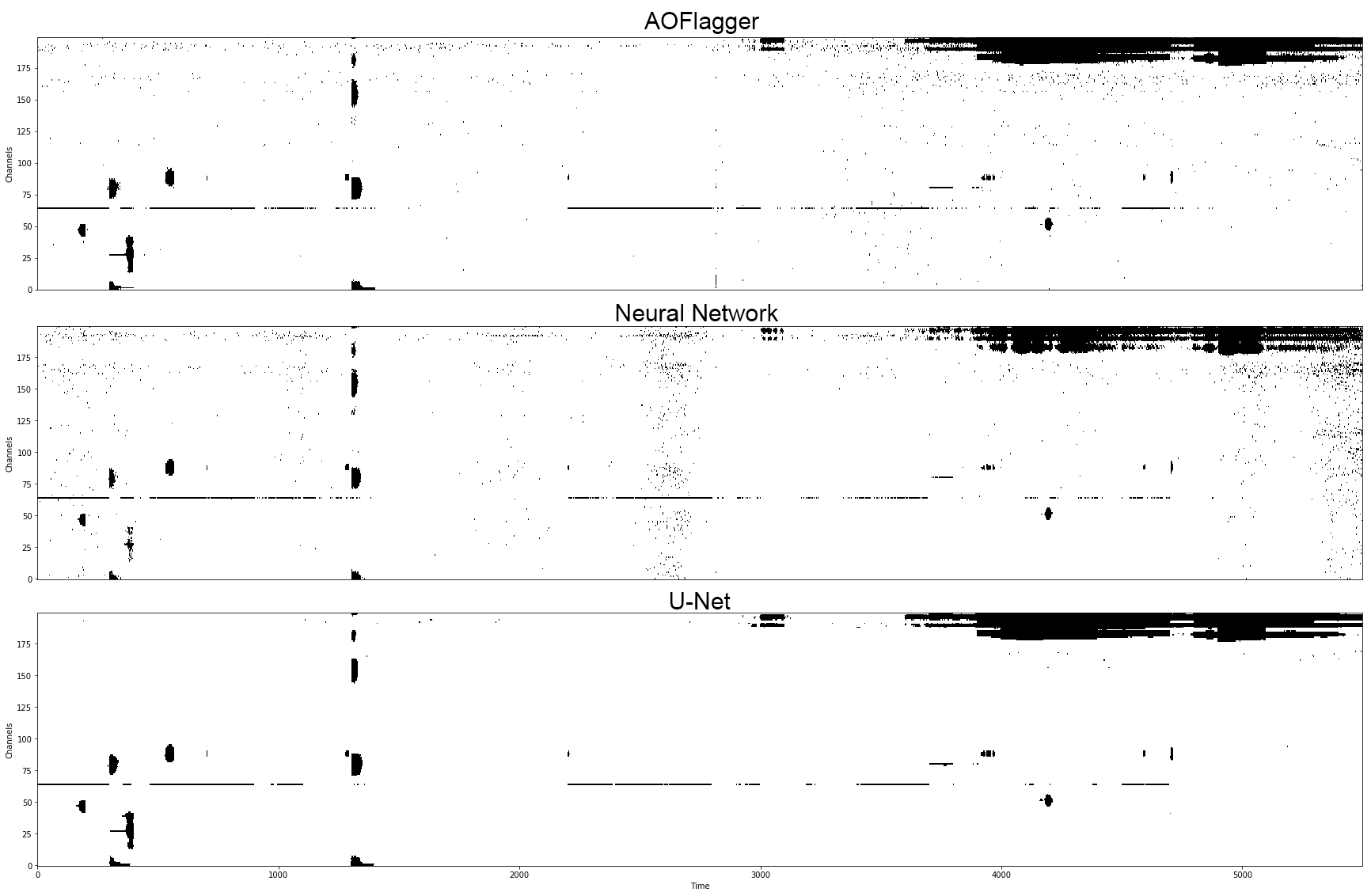}
\caption{Comparison of a single baselines Boolean flag map identifying RFI from each algorithm. It can be seen the NN method detects more RFI single sample blips that its training data. It is unknown whether these may represent more accurate results but they present false-positives when comparing evaluation metrics. The U-Net algorithm presents with almost no blips as it is likely favouring classification based on nearby classes. As the NN treats each sample independently without knowledge of nearby RFI the fact that blips across all channels at around 2800s are detected is promising.}
\label{fig:flags}
\end{figure}

An example of the flagging difference between the different methods is shown in Figure \ref{fig:flags}. While these results visualised across multiple baselines show many similarities, it can be seen that the U-Net approach struggles to identify the sporadic blips - while the NN approach appears to identify additional and different blips to AOFlagger.

Table \ref{tbl:pybdsf} shows a comparison in evaluation metrics between the images derived from each technique. It can be seen that U-Net is the only implementation which results in a positive background mean. This is likely due to the increase in flux density from PyBDSF fitting less Gaussians during processing than it did to the others.

\begin{table}[h]
\caption{Comparisons of PyBDSF Metrics}
\label{tbl:pybdsf}
\begin{tabular}{|l|l|l|l|}
\hline
 & \textbf{AOFlagger} & \textbf{NN} & \textbf{U-Net} \\ \hline
\textbf{Bg. mean (Jy/beam)} & -6e-05 & -1.3e-05 & 1.1e-04 \\ \hline
\textbf{Bg. rms (Jy/beam)} & 1.54e-3 & 1.65e-3 & 1.54e-3 \\ \hline
\textbf{Flux Density (Jy)} & 5.142 & 5.345 & 5.301 \\ \hline
\textbf{Source Count} & 59 & 65 & 64 \\ \hline
\end{tabular}
\end{table}

\section{CONCLUSIONS}
The evaluation metrics in Table \ref{tbl:metrics} show a high predictive capability for the U-Net despite zero padding influencing 4 pixels of each image. The high accuracy in U-Nets predictive results are likely causing the training of false-positives to reoccur in post-training dataset prediction. This overfitting is not an ideal result. As the goal of this work was to not only show how existing flagging strategies can be learnt by neural network implementations, but to attempt to reduce false-positive RFI predictions in the ground truth.

The increase in image source counts for the two neural network implementations imply a more precise Boolean flag map has been obtained compared to the existing algorithm, this is possibly due to the implementation of AOFlagger flagging excess amounts of useful data which the implementations do not. As described in Figure \ref{fig:flags} the NN showed an improved detection of intermittent transient RFI, which is notoriously difficult to identify.  Without simulated data a ground truth is unknown so any deductions made are purely speculative. Yet the results show how localised intermittent RFI was identified, where no spacial relationship exists as each time/frequency sample is treated independently. This is promising evidence of the network learning high dimensional features which work to discriminate RFI in a different manner to traditional thresholding techniques. 

Concrete validation of results from machine learning and neural network methods are difficult to conclude and far more testing is required on varied datasets. Yet these results show how an implementation of a simple single layer fully connected neural network are comparative to complex convolutional architectures.

\section{ACKNOWLEDGMENT}
We thank IDIA for the opportunity to pursue this work, without their fellowship and none of this would have been possible. Their cloud compute platform and storage capacity proved invaluable.

% References should be produced using the bibtex program from suitable
% BiBTeX files (here: strings, refs, manuals). The IEEEbib.bst bibliography
% style file from IEEE produces unsorted bibliography list.
% -------------------------------------------------------------------------
\bibliographystyle{IEEEbib}
\bibliography{references}

\begin{thebibliography}{10}

\bibitem{ref_ant}
Cecilia Barnbaum and Richard Bradley,
\newblock ``A new approach to interference excision in radio astronomy:
  Real-time adaptive cancellation,''
\newblock {\em The Astronomical Journal}, vol. 116, pp. 2598, 12 2007.

\bibitem{blanking}
Noppasin Niamsuwan, Joel Johnson, and Steven W.~Ellingson,
\newblock ``Examination of a simple pulse-blanking technique for radio
  frequency interference mitigation,''
\newblock {\em Radio Science - RADIO SCI}, vol. 40, 10 2005.

\bibitem{offringa_soft}
{Andr{\'e}} Offringa,
\newblock ``Aoflagger,'' \url{https://sourceforge.net/p/aoflagger/wiki/Home/}.

\bibitem{aoflagger}
Yvette Cendes and et~al,
\newblock ``Rfi flagging implications for short-duration transients,''
\newblock {\em Astronomy and Computing}, vol. 23, 04 2018.

\bibitem{offringa}
{Andr{\'e}} Offringa,
\newblock {\em Algorithms for radio interference detection and removal},
\newblock Ph.D. thesis, 2012,
\newblock Relation: https://www.rug.nl/ Rights: University of Groningen.

\bibitem{casa}
NRAO,
\newblock ``Casa guides,''
  \url{https://casaguides.nrao.edu/index.php/Main_Page}.

\bibitem{tfcrop}
NRAO,
\newblock ``Tfcrop,''
  \url{https://casa.nrao.edu/Release3.4.0/docs/userman/UserMansu161.html}.

\bibitem{uv}
Srikrishna Sekhar and Ramana Athreya,
\newblock ``Two procedures to flag radio frequency interference in the uv
  plane,''
\newblock {\em The Astronomical Journal}, vol. 156, 10 2017.

\bibitem{knn}
Naomi Altman,
\newblock ``An introduction to kernel and nearest neighbor nonparametric
  regression,''
\newblock 1992.

\bibitem{rfc}
{Tin Kam Ho},
\newblock ``Random decision forests,''
\newblock in {\em Proceedings of 3rd International Conference on Document
  Analysis and Recognition}, Aug 1995, vol.~1, pp. 278--282 vol.1.

\bibitem{bayes}
Pouria Kaviani and Sunita Dhotre,
\newblock ``Short survey on naive bayes algorithm,''
\newblock {\em International Journal of Advance Research in Computer Science
  and Management}, vol. 04, 11 2017.

\bibitem{ska}
Olorato Mosiane, Nadeem Oozeer, Arun Aniyan, and Bruce Bassett,
\newblock ``Radio frequency interference detection using machine learning.,''
\newblock {\em IOP Conference Series: Materials Science and Engineering}, vol.
  198, pp. 012012, 05 2017.

\bibitem{akeret2017radio}
Joel Akeret and et. al,
\newblock ``Radio frequency interference mitigation using deep convolutional
  neural networks,''
\newblock {\em Astronomy and Computing}, vol. 18, pp. 35--39, 2017.

\bibitem{hera_sim}
Joshua Kerrigan and et~al,
\newblock ``Optimizing sparse rfi prediction using deep learning,''
\newblock {\em Monthly Notices of the Royal Astronomical Society}, Jul 2019.

\bibitem{adam}
Diederik Kingma and Jimmy Ba,
\newblock ``Adam: A method for stochastic optimization,''
\newblock {\em ICLR}.

\bibitem{unet}
Olaf Ronneberger, Philipp Fischer, and Thomas Brox,
\newblock ``U-net: Convolutional networks for biomedical image segmentation,''
\newblock 05 2015.

\end{thebibliography}

\end{document}